\documentclass[9pt,twocolumn,twoside]{pnas-new}
\templatetype{pnasbriefreport} % Choose template 
\usepackage{amsmath,breqn}
\usepackage{bbm}
\usepackage{graphicx}
\usepackage{amsfonts}
\usepackage{xcolor}
\usepackage{upgreek}

\setboolean{displaywatermark}{false}

\makeatletter
\def\captionof#1#2{{\def\@captype{#1}#2}}
\makeatother
\usepackage{hyperref}

\def\bra#1{\mathinner{\langle{#1}|}}
\def\ket#1{\mathinner{|{#1}\rangle}}

\newcommand{\tr}{\rm{tr}}

\def\one{\mathbbm{1}}
\def\bra#1{\mathinner{\langle{#1}|}}
\def\ket#1{\mathinner{|{#1}\rangle}}

\def\aa{{\rm a}}

\def\tr{{{\rm tr}}}

\def\one{\mathbbm{1}}

\title{Quantum many-body attractors}

\author[a,1]{Berislav Bu\v{c}a}
\author[b]{Archak Purkayastha}
\author[b,c]{Giacomo Guarnieri}
\author[b]{Mark T. Mitchison}
\author[a,d]{Dieter Jaksch} 
\author[b]{John Goold}

\affil[a]{Clarendon Laboratory, University of Oxford, Parks Road, Oxford OX1 3PU, United Kingdom}
\affil[b]{School of Physics, Trinity College Dublin, College Green, Dublin 2, Ireland}
\affil[c]{Dahlem Center for Complex Quantum Systems, Freie Universit\"{a}t Berlin, 14195 Berlin, Germany}
\affil[d]{Institut für Laserphysik, Universität Hamburg, 22761 Hamburg, Germany}

\leadauthor{Bu\v{c}a}

\authorcontributions{BB discovered strictly local dynamical symmetries and dynamical quantum threads, proposed the spin lace model, initiated and conceived the study. AP and BB found the analytical results and AP performed the numerical study. All authors discussed and analysed the results and their physical implications and contributed to writing the manuscript.}
\authordeclaration{The authors declare no competing interests.}
\correspondingauthor{\textsuperscript{1}To whom correspondence should be addressed. E-mail: berislav.buca@physics.ox.ac.uk}


\begin{abstract}
Complex dynamics when occurring autonomously, i.e. without external driving, is usually associated with everyday length scales and classical physics, e.g. living organisms. This dynamics is \emph{not} quantum coherent. Quantum coherent dynamics is, by contrast, assumed to be either simple periodic oscillation in particular when autonomous, e.g. spin precession, or random quantum fluctuations. Combining autonomous complex and quantum coherent dynamics on microscopic length-scales could allow for novel coherent quantum machines working without external time-dependent driving. Motivated by this, here we provide an exact theoretical condition for a system to display complex quantum coherent dynamics on both microscopic and macroscopic length scales that we call a \emph{dynamical quantum algebraic thread} (D-QAT). Due to D-QATs our autonomous quantum coherent dynamics is robust to realistic imperfections (including low-doped disorder) and present for generic initial states, allowing for potential realisations in experiments. We give an example of a \emph{spin lace} model structurally similar to magnetic azurite and certain recently experimentally realized large single-molecular magnets with long coherence times. Our work opens the possibility for many potential applications including ultra-dense storage and manipulation of quantum memories, creating \emph{giant} quantum coherent qubits, or microscopic quantum mechanism perform complicated motion. 

\end{abstract}

\dates{This manuscript was compiled on \today}
\doi{\url{www.pnas.org/cgi/doi/10.1073/pnas.XXXXXXXXXX}}

\begin{document}

\maketitle
\thispagestyle{firststyle}
\ifthenelse{\boolean{shortarticle}}{\ifthenelse{\boolean{singlecolumn}}{\abscontentformatted}{\abscontent}}{}

\dropcap{T}he presence of quantum coherence in many-body systems can influence their material and thermodynamic properties, such as the possibly coherent transport in photosynthesis \cite{photo1}. However, theoretically understood cases of such quantum coherent many-body dynamics either require (A) fine-tuning of the initial state \cite{scars}, the system and its coupling to the environment \cite{Lidar,Viola}, or (B) extremely low temperature without coupling to the environment \cite{Kitaev}. Therefore, these constitute fragile, rather than robust coherent dynamics which is rather simple (e.g. periodic oscillations at one frequency). By contrast, classical systems generically exhibit complex dynamics with certain macroscopic degrees of freedom being non-stationary and other degrees of freedom equilibrating to stationarity. This, however, happens only at large length scales and without quantum coherence (see Fig.~\ref{fig:illu}). Here we introduce a dynamical symmetry \cite{BucaTindallJaksch,Marko1,Marko2} condition called the \emph{dynamical quantum algebraic thread} (D-QAT), which guarantees coherent complex dynamics in many-body systems without any fine-tuning on both large and microscopic length scales. As an example, we focus on spin-1/2 plaquettes. Due to D-QAT, these systems may be coupled to \emph{anything} while retaining their coherent dynamics at any temperature. Each D-QAT acts on the system as an effective qubit. A system formed out of such plaquettes, that we call a \emph{spin lace}, contains an extensive number of local dynamical symmetries, rendering it quasi-integrable, and displays complex dynamics at all length scales for both microscopic and macroscopic observables. These features are robust, even to low doped disorder. In fact, finite-temperature equilibrium states are perturbatively unstable to complex dynamics, i.e. a generic perturbation will induce persistent complex dynamics in any equilibrium state in the linear response sense. This feature should be contrasted with time crystals \cite{Wilczek} that are unstable to persistent oscillations only at a single or few frequencies, and quantum many-body scars that display oscillations for only fine-tuned initial states \cite{scars}. Furthermore, we also remark on experimental ramifications for ferromagnetic materials. The superextensive (exponential) number of D-QATs in the spin lace overlap in physical space with each in its own seperate subspace and define protected localised excitations existing in the same physical space. This may allow for storing quantum information in a dense fashion.

\section*{Results and Discussion}
\begin{figure}[!]
	\begin{center}
		\vspace{0mm}
		\includegraphics[width=0.5\textwidth]{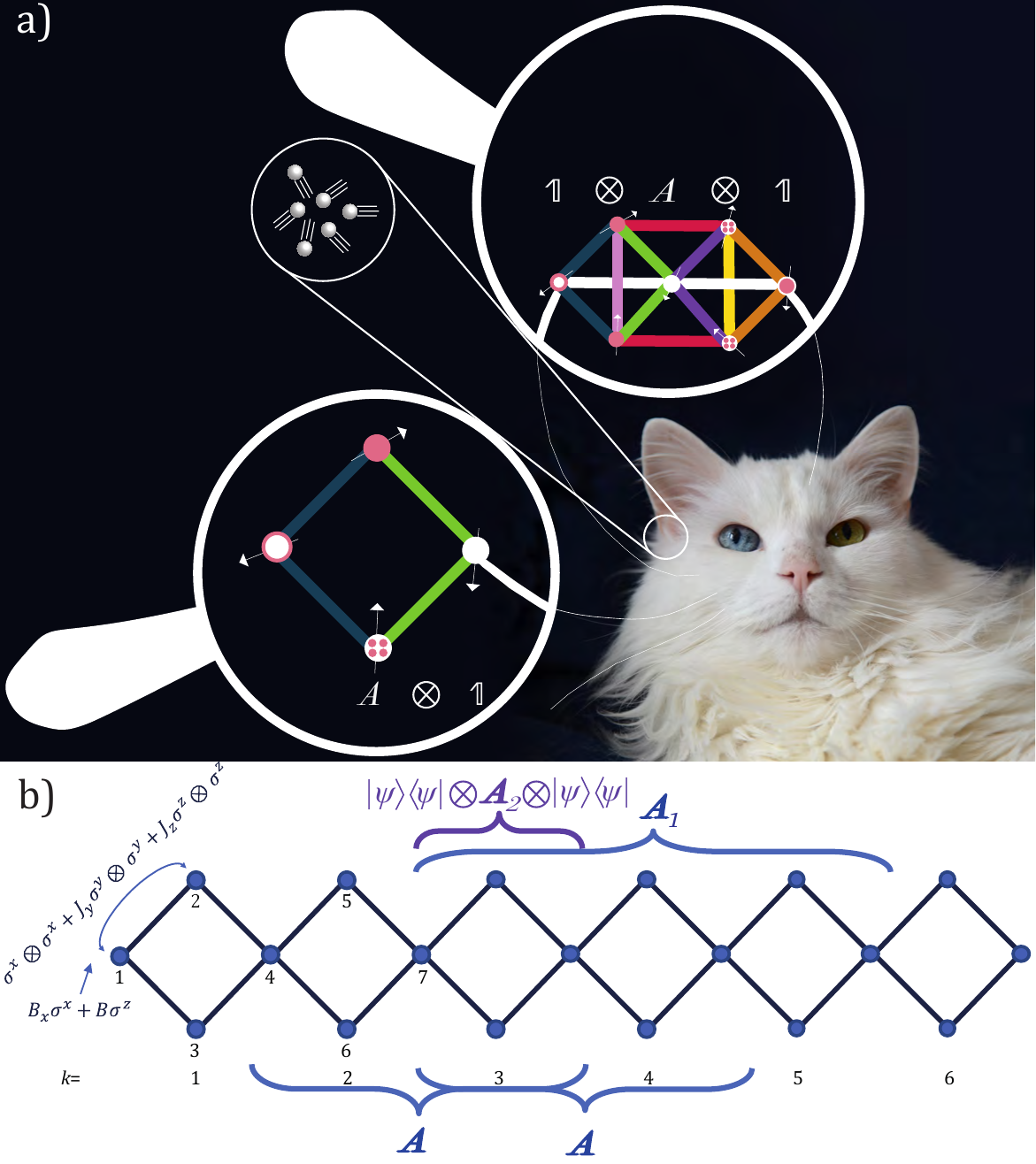}
		\vspace{-8mm}
	\end{center}
	\caption{ a) A cat is an autonomous classical system whose macroscopic center of mass degree of freedom displays complex non-stationary dynamics, whereas the microscopic degrees of freedom (motion of individual atoms) have thermalized and only show incoherent chaotic motion. Two general spin plaquettes with D-QATs are attached to the cat's whiskers. One plaquette has a  dynamical edge mode D-QAT ($A \otimes \one$), and the other a strictly local dynamical symmetry in the bulk ($\one \otimes A \otimes \one$). The plaquettes can be coupled to any other system (here a cat) on the chosen sites while retaining the D-QATs. The colors indicate the different interactions and coupling between the sites and symbols on the sites on-site potentials. b) The spin lace model from the main text. Interactions, spin sites and plaquettes ($k$) are labelled. Various examples of overlaping D-QATs are illustrated - strictly local dynamics symmetries (in blue) with $\one$ suppressed for clarity and more general D-QATs (in purple).} 
\label{fig:illu}
\end{figure}

 Consider two Hilbert spaces ${\cal H}_1$ and ${\cal{H}}_2$. The goal will be to couple the respective Hamiltonians $h_1 \in {\cal B}({\cal H}_1)$ and $h_2 \in {\cal B}({\cal H}_1 \otimes {\cal H}_2)$ such that we have a total $h=h_1\otimes \one_2+h_2$ with a D-QAT condition, i.e.,
\begin{equation}
[h,A\otimes \ket{\psi}\bra{\psi}]\ket{\psi}=\omega \left[A\otimes \ket{\psi}\bra{\psi}\right] \ket{\psi}, \label{DQAT}    
\end{equation} 
where $\ket{\psi} \in {\cal H}_2$, $A \in {\cal B}({\cal H}_1 )$.
 We also require $\tr_2(h_2)=0$ (i.e. $h_2$ acts only non-trivially on ${\cal H}_2$. Solving for $h,h_1,h_2$ for a given Hilbert space is in general intractable. Furthermore, even if one can solve for such a system there is no guarantee to find realistic Hamiltonians. 
For the moment, for sake of simplicity, we thus specialize slightly to the strictly local dynamical symmetry case of D-QAT, i.e. when \eqref{DQAT} holds for all $\ket{\psi}$. The identity then guarantees that we can embed the system Hamiltonian $h$ into \emph{any} other larger system via these sites while preserving the dynamical symmetry. In that case we say that the system has a strictly local dynamical symmetry (Fig.~\ref{fig:illu}a)). This means that local observables $O$ that have overlap with $A$, i.e. $\tr(A^\dagger O)\neq$ are expected to persistently oscillate with frequency $\omega$ \cite{Marko2}. This is true both for quenches from generic initial states and for autocorrelation functions at all temperatures. As these operators are strictly local this dynamics will be present both on microscopic and macroscopic length scales as illustrated in Fig.~\ref{fig:illu}. The approach we will take to construct models with a D-QAT is based on simple symmetry considerations. We focus on two examples. 

First take ${\cal H}_1$ to be 3 spin-$1/2$ and ${\cal H}_2$ to be a single spin-$1/2$. Upon demanding that $h_1$ is reflection symmetric around one of the sites, we find that the conserved parity-antisymmetric subspace is two-dimensional. We now assume that the total 4 spin-1/2 interaction couple these two states to the ${\cal H}_2$ spin-1/2 with equal rates. This immediately guarantees an existence of a D-QAT $A$ satisfying Eq.~\eqref{DQAT}. It is a straightforward calculation to see that any reflection symmetric interaction around sites $2$ and $3$ of ${\cal H}_1$ and the site of ${\cal H}_2$ (which we call site 4) such that site $2$ and $4$ are uncoupled fulfills this criteria (see Methods). We thus have a general class of $h$ acting on 4 site diamond plaquette (see 
~\ref{fig:illu}) with a dynamical \emph{edge mode}.  

The corresponding $A$ is calculable, but quite complicated as discussed in the SI. 
As shown in Fig.~\ref{fig:illu} any arbitrary system could be coupled to such plaquettes through chosen single spin sites while retaining the dynamical edge mode. 

Motivated by this example we also consider ${\cal H}_1$ to be a 5 spin-$1/2$ (sites 2-6) and ${\cal H}_2$ to be a 2 spin-$1/2$ (sites 1 and 7).
Now we will demand that the system has a reflection symmetry around sites 2 and 3 (given by operator $R_1$) and around 5 and 6 (given by operator $R_2$). This constrains the geometry of $h_1$ (see Fig.~\ref{fig:illu}). 

We demand that the $A$ is an identity on sites 1 and 7, and that has non-zero support in the $-1$ subspaces (reflection anti-symmetric) of $R_1$ and $R_2$, with corresponding projectors labeled as $P_{1,2}$. Similarly as before, we find a non-trivial solution for $A$ provided that sites $1$ and $7$ are not coupled to site $3$; even though one could find solutions also in that case, which would contain 3 and higher body operators, we do not focus on them (see SI). 

Note that, in general, the models do not have any simple continuous symmetries such as $SU(2)$. 
From now we specialize to a simpler case. We can now build larger models with the 4-site plaquettes as a unit cell. In particular we may build a quasi-one-dimensional spin lace model, as shown in Fig.~\ref{fig:illu}. Such a system will possess strictly local dynamical symmetries at each plaquette and more general D-QATs. 

These models show a rich and complex dynamics; to facilitate easier numerical study we will only focus on a straightforward physical example of one such row of the spin lace. Let $\sigma_x^\alpha$ be the Pauli matrices on site $x$. We label $N$ sites from $-N/2$ to $N/2$ and couple $(N-4)/3$ 4-site plaquettes $h^{(k)}, k=0,1,2,\ldots$ into a spin lace model as illustrated in Fig.~\ref{fig:illu} and we take,
\begin{dmath}
\chi_{j,k}=J_x \sigma^x_{j} \sigma^x_{k}+J_y \sigma^y_j \sigma^y_{k}+J_z \sigma^z_j \sigma^z_{k}+B (\sigma^z_j+\sigma^z_{k})+B_x (\sigma^x_j+\sigma^x_{k}),
\end{dmath}
and,
\begin{align}
&h^{(0)}= \chi_{1,2}+\chi_{2,3}+\chi_{3,4}+\chi_{1,4}  \\
&h^{(k)}= \chi_{3k,3k+2}+\chi_{3k+2,3k+3}+\chi_{3k+3,3k+4}+\chi_{3k,3k+4}\nonumber
\end{align}
which we couple in the full Hamiltonian simply as $H=\sum_k h^{(k)}$. Let $A_k=P_{k-1}\otimes\ket{\psi}\bra{\phi}\otimes P_k$, where $P_k$ is again the projector to the $-1$ subspaces of the reflection operator on plaquette $k=1,\ldots,(N-4)/6$. The $A_k$ on the double-plaquettes are given by,
\begin{align}
&\ket{\psi}=\ket{0}+b\ket{1}\nonumber\\
&\ket{\phi}=\ket{0}+b^{-1}\ket{1},
    \end{align}
with,
\begin{equation}
b=\frac{-2B+\omega}{2 B_x}.
\end{equation}
For this choice we have that frequency is just the total magnetic field, 
\begin{equation}
\omega=2\sqrt{B^2+B_x^2}. 
\end{equation}

Note that if $A_k$ is a strictly local dynamical symmetry, then $[H,A_k^\dagger A_k]=[H,A_k A_k^\dagger]=0$. The spin lace will have an extensive number of strictly local symmetries. This renders it quasi-integrable. Furthermore, it may be understood as a natural case of many-body localization in a translationally invariant system \cite{Dima}. This makes it very interesting in its own right, but we will study it here only as an example of our more general framework. We emphasize that the case we study here is special, a general spin lace model with the symmetry structure discussed before will have D-QATs. Furthermore, it will have an extensive number of them even for below-maximum-doping disorder. By that we mean that there is a random disorder on any site $j$ with some probability $p<1$. This type of disorder is natural from the perspective of doping with impurities. 

The spin lace will have further frequencies coming into the dynamics due to additional D-QATs. These are most transparent when looking at $U(1)$ symmetric version of the model $J_x=J_y$, $B_x=0$. For that case we may take in Eq.~\eqref{DQAT} $\ket{\psi}=\ket{0},\ket{1}$ and many more solutions for $A$ appear with distinct frequencies (see Methods). These frequencies, unlike generic many-body dynamics \cite{eigenstatedephasing} do not dephase even for generic initial states because there are a superextensive (exponential) number of such D-QATs overlapping on the same physical spin sites. The contrast with similar algebras for quantum many-body scars (first studied for open quantum scars in the Supp. Mat. of \cite{BucaTindallJaksch}, and later in e.g. \cite{scardynsym2}) is now clear: the latter only display persistent oscillations for fine-tuned initial states, whereas for systems with D-QAT a multitude of non-dephasing frequencies will occur at any temperature and for generic initial states (cf. related models in \cite{Arnab}.

\begin{figure}
\includegraphics[width=\columnwidth]{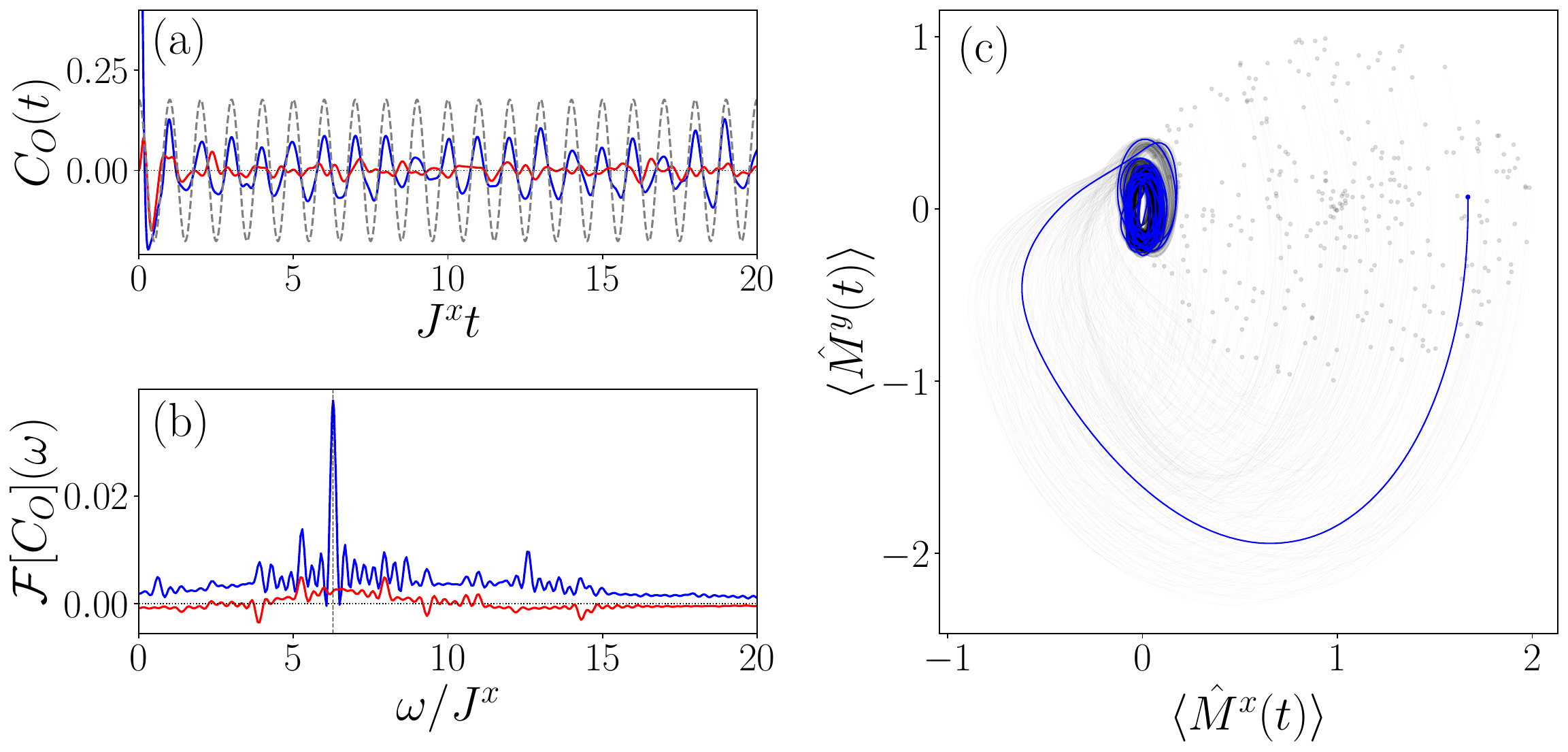} 
\caption{ (a)~Infinite-temperature correlation functions $C_O(t) = \langle O(t)\sigma_{3(j-1)+3}^x\rangle$ for chosen microscopic observables, describing the linear response of the observable $O(t)$ to a local perturbation applied to one lattice site. Persistent oscillations are seen for observables which overlap with a dynamical symmetry operator, such as $\sigma_r^x$ (blue solid lines) or $A_k$ (gray dotted line) itself. In contrast, the response of non-conserved observables, e.g.~the total spin operator $S_j^x=\sigma^x_{3(j-1)+2}+\sigma^x_{3(j-1)+3}$ (red solid lines), decays to small fluctuations around zero, showing complete loss of memory of initial conditions. (b)~Fourier transforms $\mathcal{F}[C_O(t)](\omega)$ of the correlation functions shown in~(a). The dominant frequency $\omega=2B$ (vertical dashed line) is clear in the response of $\sigma^x_{3j-2}$, while the response of $S^x_r$ shows no special feature at that frequency.  (c)~Attractor-like dynamics of the total macroscopic magnetisation vector ${\hat{M}^\alpha = L^{-1}\sum_r\left(\sigma_{3j-2}^\alpha + S_r^\alpha\right)}$ in the $x$-$y$ plane, where $L$ is the number of lattice sites. Each line shows an evolution that starts from a different initial condition and is drawn over time into a complex orbit within the attractor-like region. The attractor is visible as the shaded ellipse where all spin trajectories converge. The blue line highlights one representative trajectory. The initial conditions are generated by preparing the system in an infinite-temperature state and then projecting the total spin state of all unit cells in a given direction. See the SI for details. Parameters: $J^x=1,J^y=2,J^z=0.5, B=\pi$.} 
\label{fig:corrs}
\end{figure}

In order to demonstrate non-stationarity and coherence at arbitrary temperature we study infinite temperature autocorrelation functions as shown in Fig.~\ref{fig:corrs}. We see the emergence of complex dynamics for observables that have overlap with $A$ for both microscopic and macroscopic observables and the co-existence with thermalization as promised in Fig.~\ref{fig:illu}. 
The spin plaquette geometries that we study are very similar to existing magnetic materials with tetramer unit cells \cite{tetra1} and organic compounds much studied for their interesting mostly-low temperature properties e.g. \cite{Bogani}. The presence of D-QATs should have profound impact on the dynamics of observables even at very high temperatures. This also hints that these materials have potential for coherent quantum information processing.   

In conclusion, we have proposed dynamical quantum algebraic threads (D-QAT) generalising dynamical symmetries \cite{BucaTindallJaksch} and restricted quantum scars versions \cite{scardynsym2}. D-QATs guarantee that a quantum system may retain quantum coherent dynamics when coupled to arbitrary systems on the chosen sites. Moreover, unlike weak ergodicity breaking due to quantum scars, for systems with D-QATs quantum coherent dynamics happens even at relatively high temperature. Observables that have overlap with the D-QATs will display long-time \emph{complex} dynamics at both microscopic and macroscopic length scales. The complexity is due to the multitude of the frequencies of oscillations and the locality of D-QAT. The equilibrium states of systems with D-QATs will be unstable to complex dynamics and they genuinely interpolate between time crystals \cite{Wilczek} and complex every-day classical dynamics.  We gave a simple example of a spin lace, similar to quasi-one-dimensional magnetic materials and molecules. D-QATs of arbitrary physical size can be found and induce corresponding localized excitations with two coherent degrees of freedom. Therefore such a system could host giant qubits stable to magnetic noise and many-body fluctuations of the spin/electron dynamics with their size in real materials limited by phonons and nuclear dephasing. Moreover, the D-QATs are protected through global symmetries and are stable to even  disorder that is below maximum in doping. The dynamics of microscopic observables is partially thermalizing and partially quantum coherent non-stationary in analogy with classical complex dynamics.  

There are numerous avenues for future research, both mathematical and applicable. We only give a few examples for the concrete single example of the spin lace model introduced here. The spin lace has coherence that is protected at arbitrary temperature, in contrast to standard stabilizer codes that are fragile at any non-zero temperature and usually are found only in fine-tuned models requiring manipulation of non-local degrees of freedom \cite{Kitaev} and this can be exploited in future quantum error-correcting schemes. Any possible relation with lattice gauge theories should be explored \cite{Dima}. In that context the demand of strictly local dynamical symmetries could be relaxed and we can embed Hamiltonians with extensive dynamical symmetries in the $Z_2$ subspaces. Apart from the single exemplary spin-lace model given here, D-QATs can be used to find models with even more general, and perhaps before unimaginable, kinds of quantum autonomous complex dynamics. 

\matmethods{

\subsection*{Additional D-QATs in the $U(1)$ spin lace model}

Here we discuss how to construct a superextensive number of D-QATs focusing on the simpler $U(1)$ symmetric case of the spin lace model. 

We start with the relation \eqref{DQAT}. Consider the most general ansatz for a one-particle (or magnon) state,
$\ket{\psi}=\sum_j a_j \ket{j}$,
where $j$ means there is a spin-up/down or particle on site $j$. Any number of adjecent plaquettes described by Hamiltonian $h$ in the spin lace is reflection symmetric, i.e. $[H,R]=0, R^2=1$. We clearly have,
\begin{align}
h \ket{\psi_n}=E_n \ket{\psi_n},
RHR R\ket{\psi_n}=E_n R\ket{\psi_n}=H R\ket{\psi_n}=E_n R\ket{\psi_n}.    
\end{align}
Thus the $\ket{\phi_n}=\ket{\psi_n}-R\ket{\psi_n}$ is an eigenstate of $h$ with nodes on the leftmost and rightmost sites. A pair of such eigenstates forms the $A=A_{nm}=\ket{\phi_n}\bra{\phi_m}$ from \eqref{DQAT} with $\ket{\psi}=\ket{0}$ or $\ket{\psi}=\ket{1}$. Due to the $U(1)$ symmetry the D-QATs on plaquette $k$ can act on the simple product eigenstates $\ket{\psi_{0}}=\ket{0\ldots0}$ (and likewise for $1$) and make an exponential number of eigenstates at equally spaced energies. 
}

\showmatmethods{} 

\acknow{BB warmly acknowledges V.~Juki\' c Bu\v ca for the name ``spin lace'' and help with Fig 1a containing Pulci. BB thanks M.~Medenjak for useful discussions and collaboration on related work, H. Katsura for critical reading and valuable corrections and  O.~Castro Alvaredo, B.~Doyon, J.~Marino, C.~Navarrete-Benlloch, A.~Polkovnikov, F.~Pollmann, and T.~Prosen for useful discussions. We acknowledge support from the European Research Council Starting Grant ODYSSEY (G. A. 758403). JG would like to thank A.~Silva and M.~Dalmonte for discussions. BB and DJ acknowledge funding from EPSRC programme grant EP/P009565/1, EPSRC National Quantum Technology Hub in Networked Quantum Information Technology (EP/M013243/1), and the European Research Council under the European Union's Seventh Framework Programme (FP7/2007-2013)/ERC Grant Agreement no. 319286, Q-MAC and DJ is supported by the Cluster of Excellence 'CUI: Advanced Imaging of Matter' of the Deutsche Forschungsgemeinschaft (DFG) - EXC 2056 - project ID 390715994. JG is supported by a SFI-Royal Society University Research Fellowship. AP acknowledges funding from European Union’s Horizon 2020 research and innovation programme under the Marie Sklodowska-Curie grant agreement No. 890884. AP acknowledges Irish Centre for High End Computing (ICHEC) for the provision of computational facilities.}

\showacknow{} 

% Bibliography
\bibliography{pnas-sample}

\end{document}